\documentstyle[float,epsfig]{article}
\begin{document}
	
\def\gC{\mbox{\boldmath $C$}}
\def\gZ{\mbox{\boldmath $Z$}}
\def\gR{\mbox{\boldmath $R$}}
\def\gN{\mbox{\boldmath $N$}}
\def\ua{\uparrow}
\def\da{\downarrow}
\def\a{\alpha}
\def\b{\beta}
\def\g{\gamma}
\def\G{\Gamma}
\def\d{\delta}
\def\D{\Delta}
\def\e{\epsilon}
\def\z{\zeta}
\def\h{\eta}
\def\th{\theta}
\def\k{\kappa}
\def\l{\lambda}
\def\L{\Lambda}
\def\m{\mu}
\def\n{\nu}
\def\x{\xi}
\def\X{\Xi}
\def\p{\pi}
\def\P{\Pi}
\def\r{\rho}
\def\s{\sigma}
\def\S{\Sigma}
\def\t{\tau}
\def\f{\phi}
\def\vf{\varphi}
\def\F{\Phi}
\def\c{\chi}
\def\w{\omega}
\def\W{\Omega}
\def\Q{\Psi}
\def\q{\psi}
\def\de{\partial}
\def\inf{\infty}
\def\ra{\rightarrow}
\def\bra{\langle}
\def\ket{\rangle}

\pagestyle{empty}
\setlength{\textwidth}{5in}
\setlength{\textheight}{7.8in}

\title{{\normalsize\bf CANONICAL TRANSFORMATION  OF THE HUBBARD MODEL 
AND  W=0 PAIRING: COMPARISON WITH EXACT DIAGONALIZATION RESULTS 
}}

\author{
{\normalsize MICHELE CINI,} 
{\normalsize ADALBERTO BALZAROTTI }\\
{\normalsize RAFFAELLA BRUNETTI, }
{\normalsize MARIA GIMELLI }
{\normalsize and } 
{\normalsize GIANLUCA STEFANUCCI}\\
{\footnotesize Istituto Nazionale di Fisica della Materia and Dipartimento di Fisica,Universit\`{a} di Roma Tor Vergata,}\\
{\footnotesize Via della Ricerca Scientifica 1- 00133 Roma, Italy}\bigskip \\ 
{\normalsize Received (2000) }\\
{\footnotesize  \vspace{-\baselineskip} }} 

\date{}

\maketitle
\thispagestyle{empty}

\renewcommand{\abstractname}{}
\renewcommand{\baselinestretch}{1.04}
\begin{abstract}
\noindent
We have recently developed a canonical transformation of the Hubbard and related models, 
valid for systems of arbitrary size and for the full plane; this is  particularly suited 
to study hole pairing.  In this work we show that exact diagonalization results of the one 
 band Hubbard model for small clusters with periodic boundary conditions  
agree well with the analytical ones obtained by means of our canonical transformation.  
In the presence of a pairing instability, the analytic approach allows us  to identify 
the Cooper pairs. They are W=0 pairs, that is, singlet two-hole eigenstates of the 
Hubbard Hamiltonian with vanishing on-site repulsion. Indeed, we find  that the Coulomb 
interaction effects on W=0 pairs are dynamically small, and  repulsive or attractive, 
depending on the filling.
\end{abstract}

\bigskip

\renewcommand{\baselinestretch}{1.08}
\noindent

\section{Introduction: the Hubbard model}

Let us consider the Hubbard model  on a square lattice of $N\times N$ 
sites, with hamiltonian
\begin{equation}
H=H_{0}+W,
\label{hamil}
\end{equation}
where
\begin{equation}
H_{0}=t\sum_{\s}\sum_{(r,r')}c^{\dag}_{r\s}c_{r'\s}
\label{hamil0}
\end{equation}
with $r,r'$ nearest neighbours and 
\begin{equation}
W=U\sum_{r}n_{r\ua}n_{r\da}.
\label{hamilw}
\end{equation}
 We use 
periodic boundary conditions and the one-body wave vectors are 
 $k=(k_{x},k_{y})=\frac{2\p}{N}(p,q)$ with $p$ and $q$ integers.

\section{W=0 pairs}

The strong on-site repulsion between two opposite spin fermions normally
prevents the formation of singlet bound states. However, the planar square  
symmetry of the  
Hubbard and related models allows singlet two body 
eigenstates of $H_{0}$  belonging to the zero  eigenvalue of $W$. We will 
refer to these states as W=0 pairs. This means that the two 
fermions of a W=0 pair do not interact directly, but only by means 
of virtual electron-hole excitations. The background particles play a crucial 
role in determining the structure of the effective interaction which 
can be, in priciple, attractive or repulsive. 

In previous works$^{1,2}$ 
we have shown how to build W=0 pairs with zero 
total momentum. There the idea was to project the determinantal state
\begin{equation}
|d(k)\ket=c^{\dag}_{k\ua}c^{\dag}_{-k\da}|vac\ket
\end{equation}
on the Irreducible Representations (irreps) of the point symmetry group 
$C_{4v}$. Remarkably, by 
projecting on the irreps $A_{2}$, $B_{1}$ and $B_{2}$ one obtains 
exclusively W=0 
pairs. Here we want to point out a new and more general criterion to get W=0 
pairs. Let ${\cal G}$ a  symmetry group
of the non interacting Hubbard Hamiltonian $H_{0}$, 
big enough to justify the degeneracy of 
the single particle energy levels. Let us consider a  
two body state of opposite spins trasforming 
as the i-th component of the irrep $\Gamma$ of ${\cal G}$:
\begin{equation}
|\Q_{i}^{(\G)}\rangle=\sum_{r_{1}r_{2}}\Q_{i}^{(\G)}(r_{1},r_{2})
    c^{\dag}_{r_{1}\ua}c^{\dag}_{r_{2}\da}|vac\rangle
\end{equation}
Then we have
\begin{equation}
 n_{r\ua}n_{r\da}|\Q_{i}^{(\G)}\rangle=
 \Q_{i}^{(\G)}(r,r)c^{\dag}_{r\ua}c^{\dag}_{r\da}|vac\rangle\equiv 
    \Q_{i}^{(\G)}(r,r)|r\ua,r\da\rangle.
\end{equation}
Let  $P_{i}^{(\G)}$ be the projection operator on the i-th 
component of the irrep $\Gamma$. Since
\begin{equation}
   P_{i}^{(\G)}\sum_{r}\Q_{i}^{(\G)}(r,r)|r\ua,r\da\rangle=
   \sum_{r}\Q_{i}^{(\G)}(r,r)|r\ua,r\da\rangle
\end{equation}
if
\begin{equation}
 P_{i}^{(\G)}|r\ua,r\da\rangle=0\;\;\;\;\;\forall r
\label{w=0cond}
\end{equation}
then
\begin{equation}
  \Q_{i}^{(\G)}(r,r)=0\;\;\;\;\;\forall r.
\end{equation}
Clearly eq.(\ref{w=0cond}) is true if and only if
\begin{displaymath}
    P_{i}^{(\G)}|r\s\rangle=0\;\;\;\;\;\forall r
\end{displaymath}
where $|r\s\rangle=c^{\dag}_{r\s}|vac\ket$. 

It is always possible to 
write $ |r\s\rangle$ as 
\begin{equation}
   |r\s\rangle=\sum_{\G\in {\cal E}}\sum_{i}c^{(\G)}_{i}(r)
    |\vf^{(\G)}_{i,\s}\rangle
    \label{rdecomp}
\end{equation}
where ${\cal E}$ is the set of the irreps of the one-body spectrum 
of $H_{0}$ and 
$ |\vf^{(\G)}_{i,\s}\rangle$ the corresponding eigenstate 
with spin $\s$.  From (\ref{rdecomp}) it follows directly that if 
$ \G'$ does not belong to ${\cal E}$
\begin{displaymath}
 P^{(\G')}|r\s\rangle=0 
\end{displaymath}
and so
\begin{displaymath}
 P^{(\G')}|r\ua,r\da\rangle=0.
\end{displaymath}
We have proven the following 

THEOREM: {\bf Let $|\Q\rangle$ be a two-body eigenstate of 
the kinetic energy $ H_{0}$ with spin $ S_{z}=0$. 
Projecting $ |\Q\rangle$ on an irrep not contained in ${\cal 
E}$, we get either zero or an eigenstate of $ H_{0}$ 
with no double occupancy.}

The singlet component of this state is a {\bf W=0 pair}, and its 
momentum does not generally vanish.

\section{Binding energy: analytic approach}

From now on we call hole the fermion created by $c^{\dag}$ and 
electron its antifermion.

The Schr\"{o}dinger equation for the ground state of our $N\times N$ 
square lattice  with $n_{h}$ holes is
\begin{equation}
H |\Psi _{0}\rangle  = E_{h}(n_{h}) |\Psi _{0}\rangle . 
\label{mb}
\end{equation}
If the non interacting ground state with $n_{h}-2$ 
holes can be written in terms of a single determinantal state, the exact 
$ |\Psi _{0}\rangle$ can always be expanded in terms of excitations over 
it: 
\begin{equation}
    |\Psi _{0}\rangle =\sum_{m}a_{m}|m\rangle +
{\sum_{\alpha }}b_{\alpha }|\alpha \rangle +{\
\sum_{\beta }}c_{\beta }|\beta \rangle +....  
\label{expansion}
\end{equation}
here $ m$ runs over pair states, $ \alpha$ 
over 4-body states ($ 2$ holes and $ 1$
electron-hole pair), $ \beta$ over 6-body ones ($ 2$ holes 
and $ 2$ electron-hole pairs) and so on. 
Eq.(\ref{expansion}) is an expansion in the number of virtual 
excitations.

In the following we set up a procedure which 1) carries out the 
``Configuration Interaction'' calculation in a compact and efficient 
way 2) separates neatly the effective interaction from the 
self-energy contributions to the ground state energy. 
To understand how this mechanism works,  for the moment we truncate the 
expansion to the $\b$ states. Then  
equation (\ref{mb}) yields
\begin{equation}
     \left( E_{m}-E_{h}(n_{h})\right) a_{m} 
+{\sum_{m^{\prime }}}a_{m^{\prime }}W_{m,m^{\prime }}
    +{\sum_{\alpha }}
b_{\alpha }W_{m,\alpha }+{\sum_{\beta }}c_{\beta }W_{m,\beta } =0  
\end{equation}
\begin{equation}
     \left( E_{\alpha }-E_{h}(n_{h})\right) b_{\alpha } 
+{\sum_{m^{\prime }}}a_{m^{\prime }}W_{\alpha,m^{\prime }}
    +{\sum_{\alpha
^{\prime }}}b_{\alpha ^{\prime }}W_{\alpha ,\alpha ^{\prime }}+{\sum_{\beta }
}c_{\beta }W_{\alpha ,\beta } =0   
\end{equation}
\begin{equation}
     \left( E_{\beta }-E_{h}(n_{h})\right) c_{\beta }
+{\sum_{m^{\prime }}}a_{m^{\prime
}}W_{\beta ,m^{\prime }}
   +{\sum_{\alpha ^{\prime }}}b_{\alpha ^{\prime
}}W_{\beta ,\alpha ^{\prime }}+{\sum_{\beta ^{\prime }}}c_{\beta^{\prime} }
W_{\beta ,\beta^{\prime}}=0.   
\end{equation}
Choosing the $ \beta$ states in such a way that 
\begin{displaymath}
    (H_{0}+W)_{\b\b'}=E'_{\b}\d_{\b\b'}
\end{displaymath}
we exactly decouple the 6-body states getting
\begin{displaymath}
   \left( E_{m}-E_{h}(n_{h})\right) a_{m}+{\sum_{m^{\prime }}}a_{m^{\prime
}}W_{m,m^{\prime }}^{\prime }+{\sum_{\alpha }}b_{\alpha }W_{m,\alpha 
}^{\prime }=0
\end{displaymath}
\begin{displaymath}
   \left( E_{\alpha }-E_{h}(n_{h})\right) b_{\alpha }+{\sum_{m^{\prime }}}a_{m^{\prime
}}W_{\alpha ,m^{\prime }}^{\prime }+{\sum_{\alpha^{\prime }}}b_{\alpha^{\prime
}}W_{\alpha ,\alpha^{\prime }}^{\prime }=0
\end{displaymath}
where $ W'$'s are the renormalized interaction coefficients. It is 
clear that if we had truncated the expansion to an arbitrary number $n$
of electron-hole virtual exitations, we should have obtained the same 
results but with further renormalizations. This is a recursion method to perform the full 
canonical transformation; it applies to all the higher order
interactions, and we can recast our problem as if only $2$ and $4$-body states
existed. Now we choose the $ \alpha$ states in such a way that
\begin{displaymath}
   (H_{0}+W^{\prime})_{\alpha,\alpha^{\prime}}=
E^{\prime}_{\alpha}\delta_{\alpha,\alpha^{\prime}} 
\end{displaymath}
getting the following eigenvalue problem
\begin{equation}
 \left( E_{h}(n_{h})-E_{m}\right) a_{m}=\sum_{m^{\prime}} a_{m^{\prime }}
\left\langle m|F[E_{h}(n_{h})]+W_{eff}[E_{h}(n_{h})]|m^{\prime }\right\rangle ,
\label{efsceq}
\end{equation}
where
\begin{equation}
    \left\langle m|F\left[ E_{h}(n_{h})\right]+W_{eff}[E_{h}(n_{h})] |m^{\prime }
\right\rangle =W_{m,m^{\prime }}^{\prime }+  
{\sum_{\alpha
}}\frac{\langle m|W^{\prime }|\alpha \rangle\langle \alpha |W^{\prime }
|m^{\prime }\rangle}{E_{h}(n_{h})-E_{\alpha }^{\prime }}.  
    \label{scatoper}
\end{equation}
Equation (\ref{efsceq}) determines the amplitudes $ a_{m}$ of 
the $ m$ states in the $ n_{h}$ hole ground state 
and the corresponding eigenvalue $E_{h}(n_{h})$ relative to the 
hole vacuum. Here $F$ is the forward scattering operator 
and $W_{eff}$ the {\bf effective interaction}. After the shift
\begin{equation}
 H_{0}\ra H_{0}-E_{h}(n_{h}-2),
\label{shift}
\end{equation}
it is perfectly consistent to interpret $ a_{m}$ as the wave 
function of the {\bf dressed pair}. It is worth to 
underline that this canonical transformation enables us to identify 
the effective interaction between two holes in a $n_{h}-2$ holes 
background self-consistently. In principle we can  work it out
analytically and the expansion is neither in $U$ nor in $t$ but in 
the number of vitual exitations. It permits us to identify the binding 
energy $|\D(n_{h})|$ too. After the shift (\ref{shift}) we write the ground state 
energy $E_{h}(n_{h})$ of the Hubbard hamiltonian as
\begin{equation}
E_{h}(n_{h})=2E_{F}+\D(n_{h})
\label{delta}
\end{equation}
where $E_{F}$ is the renormalized Fermi energy and $\D(n_{h})$ the two 
holes energy gain of the 
interacting system with respect to the non interacting one. 
If the effective interaction $W_{eff}$ is attractive and $\D(n_{h})<0$  we 
can speak of hole pairing.

\section{Pairing in the 4$\times$4 lattice Hubbard model}

In this section we want to compare the results for the binding
energy $|\D(n_{h})|$ obtained 1) from eq.(\ref{delta}) by truncating the 
canonical transformation to the $\a$ states and 2) from the definition 
\begin{equation}
\D(n_{h})=E_{h}(n_{h})+E_{h}(n_{h}-2)-2 E_{h}(n_{h}-1)
\label{delta2}
\end{equation}
by computing exactly the three ground state energies involved in 
eq.(\ref{delta2}).  The equivalence of the two definitions of the 
binding energy was shown in Refs[2],[6].

We consider a 4$\times$4 lattice Hubbard 
model with $t=-1$ and periodic boundary conditions; we investigate the 
possibility of pairing when two holes are added to a two hole 
background. The one particle energy spectrum of $H_{0}$  
has 5 equally spaced levels having degeneracy 1,4,6,4,1 respectively.

Our first task is to determine of the symmetry group ${\cal G}$. 
The Space Group containing the translations and the 8 $C_{4v}$ 
operations is not enough to explain the degeneracy 6. Indeed the 
largest dimension of the irreps of the Space Group for this 4$\times$4 
lattice is 4. The Space Group has 128 elements and 20 classes. 
As observed by  previous authors$^{3,4}$ 
there must be additional 
space symmetries. We have found how to obtain them. Label the sites 
from 1 to 16 in the natural way; then rotate the plaquettes 1,2,5,6 
and 11,12,15,16 clockwise and the other two 
counterclockwise  by 90 degrees. This is one of the missing space 
symmetries, as one can see that each site preserves its first 
neighbours.  It is not an isometry. The other missing symmetries are 
generated by adding this one to the Space Group. 
In this way we obtain the Symmetry Group ${\cal G}$ of 384 elements, 
which is large enough
 to justify the degeneracy of each one particle energy level. It 
 still has 20 classes; the dimensions of the irreps are 
 1,1,1,1,2,2,3,3,3,3,4,4,4,4,6,6,6,6,8,8.

The exact interacting ground state of the system with 4 holes is 
threefold degenerate and belongs to an irrep  of ${\cal G}$ that contains 
the irreps $A_{1}$ and $B_{1}$ of $C_{4v}$ once and two times 
respectively. This irrep is not contained in the one 
particle  spectrum of $H_{0}$, which does not admit degeneracy 3.
By the above theorem, pairs belonging to it must be W=0 pairs. These 
W=0
pairs arise from the single-particle states  of $H_{0}$ with eigenvalue 
-2, and their irrep is contained in the square of the irrep of the 
single-particle level.

The $m$ states of the expansion (\ref{expansion}) are  
all the W=0 pairs belonging to the ground state irrep.  We have computed the 
binding energy $|\D(4)|$ analitically from eq.(\ref{delta}) by truncating 
the expansion (\ref{expansion}) to the $\a$ states and also by means of 
exact diagonalization  from eq.(\ref{delta2}). The results are 
listed below for different values of the on-site interaction $U$ 
(energies are in 
eV):

\vspace*{0.5 cm}
\begin{center}
\noindent 
\begin{tabular}{|c|c|c|c|c|}
\hline 
  & $U=$0.1 & $U=$0.5 & $U=$0.7 & $U=$1  \\
\hline 
$\D(4)_{exact}$ & -0.053 & -0.32 & -0.67 & -1.18 \\
\hline 
$\D(4)_{analytic}$ & -0.078 & -1.95 & -3.83 & -7.81 \\
\hline 
\end{tabular}
\end{center}
\vspace*{0.5 cm}
\noindent
As  expected, with increasing $U$ the difference 
$|\D(4)_{exact}-\D(4)_{analytic}|$ increases because the 
renormalization induced by virtual electron-hole exitations becomes 
 important and it is no longer a good approximation 
to consider the $\a$ states only. Nevertheless the canonical 
transformation still predicts the right sign of $\D$. 

The above canonical transformation applies when two holes are added 
to a determinantal vacuum. To study the system at and close to half filling,
we have extended the above canonical transformation to the case when 
the vacuum is degenerate.  We 
are able to demonstrate  analytically that holes pair and to
explain 
 the ground state symmetries that were found numerically in 
Ref.[3],[5]. 
Thus,  W=0 pairs are responsible for pairing and for 
the symmetry of the interacting ground state. This  is also 
confirmed by the superconducting quantization of magnetic flux; in
previous works$^{2,6}$ we show  
that the $C_{4v}$ symmetry is restored exactly at half fluxon 
$\f_{0}/2=hc /2e$ allowing the existence of W=0 pairs. The 
symmetry of the interacting ground state is still the same of the possible 
W=0 pairs and the corresponding energy versus flux has there a second minimum.
A full account of the theory will be submitted elsewhere$^{7}$. It is 
clear by now, however, that the existence of bound pairs of 
nonvanihing momentum opens up the possibility of a Jahn-Teller 
distortion and the present approach is likely to predict charge
inomogeneities.

\section*{Acknowledgements}
This work has been supported by the Is\-ti\-tu\-to Na\-zio\-na\-le di Fisica
della Materia. 

\section*{References}

{1.   Michele Cini, Gianluca Stefanucci and Adalberto Balzarotti,
Solid State Communication {\bf 109}, 229 (1999).\\
\noindent
{2. Michele Cini, Gianluca Stefanucci and Adalberto Balzarotti, 
European Physical Journal {\bf B 10}, 293 (1999).\\
\noindent
3. G. Fano, F. Ortolani and A. Parola, Phys. Rev. {\bf 
B 46}, 1048 (1992).\\
\noindent
4. J. Bonca, P. Prelovsek and I. Sega, Phys. Rev.  
{\bf B 39}, 7074 (1989).\\
\noindent
5. A. Parola, S. Sorella, M. Parrinello and E. Tosatti, 
Phys. Rev. {\bf B 43}, 6190 (1991).\\
\noindent
6. Michele Cini and Adalberto Balzarotti, Phys. Rev.  
{\bf B 56}, 14711 (1997); Michele Cini Adalberto Balzarotti and 
Gianluca Stefanucci, European Physical Journal B {\bf 14}, 269 (2000).\\
\noindent
7. Michele Cini   and 
Gianluca Stefanucci, to be published.\\
\\

\end{document}